\documentstyle[12pt, aasms4, flushrt, epsfig, rotating]{article}

\def\BEL7{/bel6/scratch/gim/DAOfinal/TRUE/RVS/TABLE}
\def\lta{\mathrel{\hbox{\raise 0.6 ex \hbox{$<$}\kern
                   -1.8 ex\lower .5 ex\hbox{$\sim$}}}}
\def\gta{\mathrel{\hbox{\raise 0.6 ex \hbox{$>$}\kern
                   -1.7 ex\lower .5 ex\hbox{$\sim$}}}}
\received{}

\begin{document}

\title{The Open Cluster NGC$\,$7789: II. CCD $VI$ Photometry}
\author{Munhwan Gim, Don A. VandenBerg}
\affil{\small{Department of Physics and Astronomy, University of Victoria, \\
 Victoria, BC, V8W 3P6}
\begin{center}
Electronic mail: \bf{gim@uvastro.phys.uvic.ca, davb@uvvm.uvic.ca}
\end{center}}

\author{Peter B. Stetson, James E. Hesser}
\affil{\small{National Research Council, 
 Herzberg Institute of Astrophysics, 
 Dominion Astrophysical Observatory  
 5071 West Saanich Road, Victoria, BC, V8X 4M6}
\begin{center}
Electronic mail: \bf{firstname.lastname@hia.nrc.ca}
\end{center}}

\author{David R. Zurek\altaffilmark{1}}
\affil{\small{Department of Physics and Astronomy, University of Victoria,\\
Victoria, BC, V8W 3P6}}
\altaffiltext{1}{\small{Present address: Space Telescope Science Institute, 
3700 San Martin Drive,  Baltimore, MD 21218.   Electronic mail: \bf zurek@stsci.edu}}

\begin{abstract}

A $(V,V-I)$--diagram for the intermediate-age open cluster NGC$\,$7789
has been derived from CCD observations of more than 15,000 stars
within $\approx 18$ arcmin of the cluster center.  From the
brightest giants and blue stragglers at $V\sim 11$ to the faintest
lower main-sequence stars that were observed (at $V\sim 21$, $M_V
\sim 9$), the C-M diagram is well defined.  A prominent clump
of core helium-burning stars is evident at $V=13.0$ and the upper
end of the main sequence shows a fairly pronounced curvature to
the red, which is indicative of significant convective core
overshooting.  Indeed, comparisons with up-to-date stellar models
show that it is not possible to explain the observed morphology in
the vicinity of the turnoff unless the overshooting is quite
extensive.  Interestingly, if sufficient overshooting is assumed
in order to match the main-sequence data, it is not possible to
reproduce the cluster's extended giant branch unless the cluster 
age is at least 1.6 Gyr (assuming a metallicity in the range
$-0.2\le$ [Fe/H] $\le 0.0$).  This, in turn, requires that the
cluster have an apparent distance modulus $(m-M)_V \le 12.2$.
Thus, sometime within the past few 
hundred million years, the ignition of
helium burning in NGC$\,$7789 has switched from a quiescent to an explosive
(``flash'') phenomenon, and the length of the cluster's red-giant
branch has been steadily increasing with the passage of time since
then.  From main-sequence fits to models that have been carefully
normalized to the Sun, we infer a reddening $0.35\le E(V-I) \le
0.38$.

\end{abstract}


\section{INTRODUCTION} \label{Intro}

The intermediate-age open cluster NGC$\,$7789 [$\alpha$(1950)~=~23$^{h}$54\fm5,
$\delta$(1950)~=~56\arcdeg~27\arcmin; $l=115\fdg49$, $b=-5\fdg36$]
has been the subject of numerous observational and theoretical studies since
Burbidge \& Sandage (1958) obtained photoelectric and photographic $UBV$ photometry 
for about 700 stars within $\sim 450^{\arcsec}$ of the cluster center.  Their 
color-magnitude diagram (CMD), which reached as faint as $V = 16.3$ ($\sim2.5$
mag below the top of its main sequence), shows a well-defined and extended
red-giant branch (RGB), a prominent ``clump'' of core He-burning stars, many 
blue stragglers, and a main sequence whose top end bends significantly to the red.  
Although the foreground reddening is fairly high, with estimates ranging from 
$E(B-V) = 0.23$ (\cite{arp62}) to 0.32 (\cite{ss70}), it appears to be 
reasonably uniform across the face of the cluster, judging from the tightness of
the observed main sequence in, especially, the recent CCD CMD by Jahn et al. (1995).  
As indicated in Table~\ref{redmet}, which also summarizes the available 
$E(B-V)$ and distance determinations, the metallicity of NGC$\,$7789 appears to be
slightly less than solar.

An extensive ~proper-motion ~membership analysis of NGC$\,$7789 was carried out by  
McNamara~\&~Solomon (1981), who identified 679 probable members brighter than 
$B\approx 15.5$ ($M_{V}\sim 2.1$).
Radial-velocity measurements by Thogersen et al. (1993), Scott et al. (1995),
and Gim et al. (1998; Paper I)
have pretty well established which of the giants are cluster members.  In
addition, there have been many investigations to ascertain the membership status
of the blue straggler candidates --- using proper motions (\cite{p75};
\cite{ms81}), radial velocities (\cite{ss70}; \cite{sh84}; \cite{ds87};
\cite{mrp89}; \cite{ml94}), or polarization data (\cite{b82}).  We review the
results of all of these investigations in section \ref{CMD}.

The fact that NGC$\,$7789 is a very populous cluster gives it considerable
potential to test stellar models (simply because its principal photometric
sequences on the CMD are very well defined).  The curvature of the upper 
main-sequence and the lack of significant numbers of subgiants is suggestive of 
convective core overshooting (\cite{mp88}), though Martinez Roger et al. (1994) 
have argued in favor of classical, non-overshooting models.  
However, none of the attempts
(to date) to fit isochrones to the observed CMD are very convincing.  On the one
hand, Mazzei \& Pigatto conclude that the age of NGC$\,$7789 is 1.1 Gyr, but the 
isochrone for that age seems too bright in the vicinity of the turnoff and it 
fails to reproduce the cluster's extended RGB.  On the other hand, the 
non-overshooting isochrones employed by Martinez Roger et al.~have a morphology 
in the vicinity of the turnoff that is not at all like that observed.  Even 
though they find a similar age (1.2 Gyr), their adopted distance modulus differs
from the value assumed by Mazzei \& Pigatto by nearly 0.5 mag.

On the observational side, too, further work is warranted.  For no more than a
small fraction of the cluster has a CMD been constructed.
In the best available CMD (Jahn et al.~1995) the smaller number of stars only
allows the main sequence to be clearly identified to $V \approx 19$ ($M_V \sim
6.5$) even though the limiting magnitude reaches $V \approx 20$. 
Much more extensive photometric coverage could be expected to reveal 
more giants, perhaps extending the RGB to redder colors, and a deeper survey 
would delineate the lower main sequence to much fainter magnitudes, thereby
offering stronger constraints on stellar models and facilitating comparisons
with the CMDs of other clusters having similar metallicities.  It is for these
reasons, and to revisit the question as to whether models with, or without,
convective overshooting are better able to explain the observations, that we
have undertaken this investigation.  

Our photometry and its reduction are discussed in the section \ref{OBS}, 
the interpretation of the data using up-to-date stellar evolutionary tracks 
and isochrones is contained in section \ref{ANA}, and a brief summary
of the main results of this investigation is given in section \ref{CON}. 

\section{OBSERVATIONS and PHOTOMETRY} \label{OBS}

\subsection{Observations}

Observations for this project were made at the f/5 modified Newtonian
focus of the Dominion Astrophysical Observatory (DAO) 1.8$\,$m
Plaskett telescope, by Zurek with the Tek-2 CCD on the nights of 1993 October
10/11, 15/16, and December 16/17, and by Gim with the SITe-1 CCD on 1995
October 31/November 1, November 1/2, December 5/6, 6/7, and 8/9.
Each detector contains 1024 $\times$ 1024 pixels at a scale of 0.53 arcsec
per pixel, yielding a 9 $\times$ 9 arcmin$^{2}$ image size.  The gain and
readout noise were 4.0 e$^{-}$/ADU and 11 e$^{-}$ for the Tek-2 CCD and 4.7
e$^{-}$/ADU and 13 e$^{-}$ for the SITe-1 CCD detector.  A Johnson
$V$, $I$ filter set was used.

Figs.~\ref{pfig1a} and \ref{pfig1b} shows the area observed in this study: Gim (1998) 
provides a full journal of the cluster and standard star observations.
We used a total of 316 frames, among which 66 (33
frames each in $V$ and $I$) were exposed for 600$\,$s, 74 (37 frames each 
in $V$ and $I$) for 60$\,$s and 76 (38 frames each in $V$ and $I$) for 6$\,$s.  
An additional ten frames had 10$\,$s exposures in $V$. There was considerable
frame-to-frame overlap among adjacent fields.  The diameter of the total
cluster area covered was roughly 35 arcmin.  The full-width at
half-maximum (FWHM) of the stellar image cores ranges from 2$\farcs$2
to 3$\farcs$2 except those taken on 1995 December 7, for which the FWHM
is about 4$\farcs$0 because telescope tracking problems and strong winds
resulted in elongated star images on some frames.  Primary standard
stars were taken from Landolt's {\it UBVRI\/} standard stars (\cite{lan92} and
\cite{s97}) and also from stars in the globular cluster NGC$\,$7006 (\cite{dav92}
and \cite{s97}) and the open cluster M$\,$67 (\cite{mmj93}). 

\subsection{Photometry}

All of the CCD image frames were subjected to two preprocessing steps,
bias subtraction and flat fielding, in order to remove the instrumental
signature; this was done with IRAF\footnote{IRAF is distributed by the
National Optical Astronomy Observatories, which are operated by the
Association of Universities for Research in Astronomy, Inc., under
contract with the National Science Foundation.}.  
After the raw images from a given night 
were corrected for the floating bias level based on the overscan
region and after they were properly trimmed, about 10 bias frames were
median-combined into one bias-pattern frame, which was then subtracted
from the raw images.  During the various observing runs, mostly dusk and
dawn sky flat frames were recorded, but on one night flat-field images
were obtained by observing the undersides of clouds. 
All frames were divided by the median-combined flat frames for each filter 
on each night to remove pixel-to-pixel sensitivity variations.

All photometric reductions and analysis following the above preprocessing
were accomplished by automatic profile-fitting techniques: DAOPHOT,
ALLSTAR (\cite{s87}) and ALLFRAME (\cite{s94}) following the detailed 
explanations given in the {\it DAOPHOT II USER's Manual\/} and Stetson (1987).
The approximate fitting function for the PSF was decided automatically
for each frame as one of six different analytic models; 
tabulated sub-pixel corrections  from the analytic model to the true empirical PSF
were allowed to vary linearly with position in the frames.  
One comprehensive star list with a unique coordinate system is required 
before running ALLFRAME.  However, it is not easy to match up all star lists
from all frames, which are partially overlapped but greatly spread out in
order to have greater areal coverage. Therefore, a set of plates of NGC$\,$7789 from
the Mount Wilson 100-inch Hooker telescope measured on the DAO PDS machine 
was used to provide a single coordinate system spanning the full area of the cluster, to which
the CCD images could be individually referred.  The program DAOMASTER
was used to estimate the individual geometric transformation equations
relating the coordinate systems of the various frames to make the
composite starlist for input to ALLFRAME.
ALLFRAME derives a self-consistent set of positions and magnitudes for all
detected star-like objects in an area of sky by using the geometric and
photometric information from all images of a given field.
Aperture corrections to place the relative PSF magnitudes 
derived from a given image on a repeatable absolute scale were 
determined by synthetic-aperture measurements of 
the same stars used for defining the PSFs. We employed
apertures ranging from small (3 pixels) to large radii (25 pixels) on
copies of the frames where all stars except the selected stars had been
digitally subtracted.

\subsection{Standardisation}

We have chosen Landolt's (1992) observations  as the basis of our standard system.
However, to improve the internal homogeneity and precision of our results,
we have combined our own measurements of Landolt's stars over the course
of 15 years with his published data, producing refined magnitudes and
colors which retain the overall zero points and color scales of the
Landolt system, but with better precision and internal consistency. Two
other primary standard fields are used in our calibrations: 28 stars in
the open cluster M67 (\cite{mmj93}) and 124 stars in the  globular cluster NGC$\,$7006
(\cite{dav92}; \cite{s97}), Both photometric systems are also closely tied
to Landolt's.

Instrumental magnitudes for all detected stars were transformed to the
standard $V$ and $I$ magnitudes using the relations  \begin{equation}
\label{eq1} v = V + a_{0} + a_{1}(V-I) + a_{2}(X - 1.25) + a_{3}T,
\end{equation} \begin{equation} \label{eq2} i = I + b_{0} + b_{1}(V-I) +
b_{2}(X - 1.25) + b_{3}T, \end{equation} where $v$, $i$ are the
instrumental magnitudes in the $V$ and $I$ filters, respectively, $X$
is the airmass, and $T$ is the time of the observation (the terms in $T$
allow for the zero-order effects of any extinction variations during the
night); the $a_{i}$, $b_{i}$ are unknown transformation coefficients.  For
three photometric nights in 1995, all the above coefficients were
determined from the primary standard stars; Table \ref{coefFix} contains
our derived values for the color-dependent coefficients $a_{1}$ and $b_{1}$
and Table \ref{coefCol} for the other corresponding values. 
We defined a local standard sequence in the field of
NGC$\,$7789 based on the above transformations for the photometric nights.
These local stars were then added to the standard list to improve the
photometric tie-in among the three photometric nights, and also to
determine the photometric zero points for individual frames obtained on
non-photometric nights. 
The standard stars were mostly observed at airmasses larger than 1.2:
for most of the NGC$\,$7789 frames the airmasses were less than 1.1, though a
significant number had higher airmasses (up to 1.6).

Figure \ref{pfig2} shows the residuals (in the sense of {\it present\/} --
{\it standard\/}) in $V$ and ($V - I$) as functions of apparent magnitude
and color for the primary standard stars. The offset and the standard
error of the mean offset in $V$ and ($V-I$) based on 274 stars are 
$-$0.0024$\pm$0.0009; s.d. $=$ 0.025 and $-$0.0019$\pm$0.0014; s.d. $=$ 0.030, 
respectively.

\subsection{Comparison with other photometry for NGC$\,$7789}

In order to check the reliability of the calibration, our $V$ magnitudes for
NGC$\,$7789 stars were compared with six previous sets of photoelectric $UBV$
photometry (\cite{bs58}; \cite{j77}; \cite{b82}; \cite{c82}; \cite{jh75};
\cite{bw80}), two sets of photographic $BV$ photometry (\cite{bs58};
\cite{m80}), and CCD photometry in $BV$ by Jahn et al (1995).  The
comparisons were made only for $V$ because no $I$-band photometry has
previously been published. The differences in $V$ magnitude, in the sense of
({\it present\/} -- {\it published\/}), as functions of $V$ magnitude and
$(V-I)$ color are shown in Figs. \ref{pfig3a}, \ref{pfig3b} and
\ref{pfig3c}. Table \ref{Tcomp} shows the mean offset and slope in
the three plots.

The mean and standard deviation of the magnitude residuals indicate that
the offset between our CCD photometry and the previous photoelectric
photometry in Fig.~\ref{pfig3a} is negligible, +0.003$\,$mag (30
stars) and $-$0.006$\,$mag (18 stars), but the standard deviations, 
0.035$\,$mag and 0.033$\,$mag, are appreciable.
The comparison with previous photographic photometry 
(Fig.~\ref{pfig3b}) shows significant
offsets: $\Delta{V} = 0.053\,$mag (284 stars) and 0.018$\,$mag (62 stars)
with standard deviations of 0.045$\,$mag and 0.061$\,$mag, respectively.
In addition, a systematic trend in $V$ magnitude with the ($V - I$) colors
was found in comparison with both photoelectric and photographic
photometry in the sense that our CCD photometry is brighter than published
photometry for the bluer stars, with a slope of 0.032 to 0.068.  In
contrast, a comparison (Fig.~\ref{pfig3c}) with other CCD photometry 
(\cite{jkr95}) appears to reveal a slight zero point shift of 
about 0.021 mag, but no systematic trend with color.

\section{The CMD} \label{CMD}

Figure~\ref{pfig4} shows the CMD for all 15617 stars with $< 0.1$ mag errors
in the measured colors that were detected at least twice on both the $V$ and
$I$ frames\footnote{The photometry files are available in machine-readable format from
the Canadian Astrophysical Data Center, Dominion Astrophysical Observatory. They
are also available from J.-C. Mermilliod, who has added our observations into 
his open cluster data bank.}. 
Photometry was obtained for 5915 additional stars, but they are
not used in the following analysis because of their relatively large errors
($\sigma(V-I)\ge 0.1$ mag) or because they had only one detection in either $V$
or $I$.  The stars rejected by one of these criteria  
are plotted in Figure~\ref{pfig5}, along with the cluster's main sequence 
fiducial fainter than $V=14.0$.  The latter is an eye-estimated, hand-drawn fit 
to the tight, well-defined lower main sequence population of NGC$\,$7789 (see Fig.~\ref{pfig4}).
When compared with Figure~\ref{pfig4}, Figure~\ref{pfig5} indicates that our 
rejection criteria have not eliminated very many probable cluster members.
Figure~\ref{pfig6} shows the standard errors in $V$ as function of $V$ magnitude. 

Our photometry covers a much larger area and extends to much fainter magnitudes
than any previous survey.  The CMD that we have obtained shows an
extended giant branch, a resonably well-defined ``clump'',
numerous blue straggler candidates (see below), and a main sequence that extends
down to at least $V=21$.  There is no evidence of any gap in the vicinity of the
turnoff, though the upper end of the main sequence bends well to the red.  This
is reminiscent of the CMDs recently obtained for e.g., NGC$\,$2420 (\cite{akst90})
and NGC$\,$752 (\cite{dlmt94}), and indicates substantial convective core
overshooting in the main-sequence phase (see, e.g., \cite{dsg94}).  Field-star
contamination, particularly in the color range $1.0 < V-I < 1.5$, is quite
severe.

Figures~\ref{pfig7a} and \ref{pfig7b} summarize what we know about the
cluster membership. 
According to McNamara \& Solomon (1981), those stars which are plotted as  
{\it plus signs} have a proper-motion membership probability $P(\mu) \ge 80$\%.  
Based on radial-velocity measurements made between 1979 and 1996, Gim et al. (1998) have 
identified 78 giant-star members: these are plotted as {\it open circles}, or
denoted by the letter ``V'' to indicate that they are possible radial-velocity
variables.  Concerning the blue straggler candidates, the evidence seems to be
quite strong that at least 16 of them (the {\it open squares} in Fig.~\ref{pfig7a} 
and \ref{pfig7b}) are
members, according to our assessment (cf. Table \ref{BSMem}) of the proper
motion, radial velocity, photometric, and polarization work 
that has been done to date to determine the membership of 48 apparent blue 
stragglers in NGC$\,$7789.  As indicated in the final column of this table, the 
membership status of 13 of these stars was considered to be uncertain, so the 
actual number of {\it bone fide} blue stragglers could well be significantly 
greater than our estimate.  Finally, it is worth noting that Jahn et al. (1995)
found one pulsating blue straggler and 15 short-period variables as the result of
observing NGC$\,$7789 over four consecutive nights. 

\section{Analysis of the NGC$\,$7789 CMD} \label{ANA}

\subsection{Comparison with the M67 CMD}
  
As indicated in Table~\ref{redmet}, the available metallicity estimates for 
NGC$\,$7789 range 
from [Fe/H] $\approx -0.3$ to 0.0 (solar), whether determined spectroscopically
or photometrically.  As the uncertainty in this quantity is quite large, making
it difficult to choose which models should be compared with the photometry, it
is clearly desirable to try to constrain this parameter more tightly.  One
way to do this is to compare, in an age- and reddening-independent way, our CMD
for NGC$\,$7789 with that of another open cluster having a similar metallicity.
M$\,$67 is an obvious choice for such a comparison cluster because it has a 
well-defined CMD on the ($V,V-I$)--plane (\cite{mmj93}) and because its metal
abundance is very well determined at [Fe/H] $\approx -0.05$: most recent
determinations  (e.g., \cite{ntc87}; \cite{ht91}; \cite{fj93}) are within 
$\pm 0.05$ dex of this estimate.  

To carry through the analysis, let us assume that the stars in NGC$\,$7789 have
exactly the same chemical composition as those in M$\,$67.  If that is the case,
and if we superpose the cluster CMDs in such a way as to force the core 
He-burning ``clump'' stars in both clusters to have the same mean magnitude and 
color, which one might naively expect, then their respective lower main sequence
populations should also overlay one another.  As shown in Figure~\ref{pfig8},
such a coincidence is not obtained: the lower main sequence of NGC$\,$7789 is 
redder than that of M$\,$67 (at a fixed magnitude), which indicates that either
our assumption of a common metallicity is incorrect, or the clump stars in
NGC$\,$7789 have not been properly fitted to those in M$\,$67.  The two clusters
have quite different ages and, according to the predictions of stellar models, 
the NGC$\,$7789 clump should be somewhat bluer and brighter than M$\,$67's, if 
the stars in both clusters undergo comparable amounts of mass loss prior to the
core He-burning phase\footnote{As shown by e.g., Demarque \& Hirshfeld
(1975), the zero-age 
horizontal branch (ZAHB) for stars more massive than $\approx$ 1.0$\,{{\cal M}_\odot}$, 
runs from the lower right to the upper left on the C-M plane
(in the direction of increasing mass).}.  Our ZAHB for solar abundances (see
below) indicates, for instance, that a 1.6$\,{{\cal M}_\odot}$ model is 0.17 mag
brighter and 0.04 mag bluer (in $V-I$), than one for 1.2$\,{{\cal M}_\odot}$.

Suppose we adopt these offsets in positioning the NGC$\,$7789 clump relative to
that of M$\,$67 and then intercompare the lower main sequence fiducials of the
two clusters.  The result is Figure~\ref{pfig9}, and once again we find that  
the unevolved stars in NGC$\,$7789 are redder than those of M$\,$67 (at a fixed 
$V>15$).  [The slope of the theoretical ZAHB is predicted to be nearly the same
as the slope of the lower main sequence; consequently, the change in $V$ 
largely compensates for the change in $V-I$ (of the clump stars) and the 
relative main-sequence locations of the two clusters are not altered 
appreciably.]  Thus, the clump stars in NGC$\,$7789 {\it must} be significantly
bluer and/or {\it fainter} than those in M$\,$67.  If the reverse were true, then the
main sequence of NGC$\,$7789 would be well separated from (and redder than) that
of M$\,$67, which would imply that the former has much greater than solar 
abundances of the chemical elements --- something for which there is no 
observational support.  

According to, e.g., Friel \& Janes (1993) and Twarog et al. (1997), NGC$\,$7789 is 
approximately 0.1 dex more metal-poor than M$\,$67, which would suggest 
that NGC$\,$7789 has [Fe/H] $\gta -0.2$.  However, the uncertainties are such 
that both clusters could have [Fe/H] $\approx 0.0$, and since it is useful to explore 
how the interpretation of a given CMD depends on the assumed metallicity, 
models for both [Fe/H]~$=0.0$ and $-0.2$ will be compared with the observations.  

\subsection {Stellar Model Fits}

For their study of NGC 6819, Rosvick \& VandenBerg (1998) computed several
sets of models for solar abundances: reference should be made to that study for
a description of the evolutionary code that was used.  Worth emphasizing is the
fact that those models were carefully normalized to the Sun and that convective
overshooting has been treated using a parameterized version of the Roxburgh (1989)
criterion.  To be specific, we have determined the radius of a convective core,
$r_{cc}$, from the requirement that
$$\int\limits_0^{r_1} {{\cal F}_{\rm over}}(L_{\rm rad}-L){1\over T^2}
{{d\,T}\over{d\,r}}d\,r + \int\limits_{r_1}^{r_{cc}} (2-{{\cal F}_{\rm over}})
(L_{\rm rad}-L){1\over T^2}{{d\,T}\over{d\,r}}d\,r = 0~,$$
where ${{\cal F}_{\rm over}}$ is a parameter with allowed values in the
range $0.0<{{\cal F}_{\rm over}}\le 1.0$, 
$L$ is the total luminosity produced by nuclear reactions, $L_{\rm rad}$ 
is the radiative luminosity, and the other symbols have their usual 
meanings.\footnote{
In its most general form, Roxburgh's 
(1989) criterion contains two integrals with limits from $r=0$ to $r=r_{cc}$ ---
one which is equivalent to the above with ${{\cal F}_{\rm over}} = 1.0$, and the 
other which properly accounts for energy dissipation.  Where the difference
between these two integrals vanishes is the correct radiative-convective
boundary.  However, the second integral requires the solution of the full set
of turbulent equations that applies to a convective core and, consequently, it
is not easily evaluated (though see \cite{ccd98}).  The parameterized
criterion that we have solved is, in contrast, extremely easy to evaluate and,
although the ``trial and error'' approach must be used to determine which value
of ${{\cal F}_{\rm over}}$ produces the most realistic models to fit to an
observed CMD, the computational effort is still rather small.  Although our
{\it ad hoc} way of limiting the extent of convective overshooting does not add
very much to the physics of convection, an evaluation of ${{\cal F}_{\rm over}}$
does provide some empirical information on the relative importance of the
dissipation term.}
The radius $r_1$ is the classical radiative-convective boundary; i.e., the point
where $\nabla_{\rm ad} = \nabla_{\rm rad}$ (and $L=L_{\rm rad}$).  Because $L<
L_{\rm rad}$ in the overshooting region ($r_1<r\le  r_{cc}$) {\it and} because
the dissipation is always positive, consistency demands that the factor 
${{\cal F}_{\rm over}}$ in the first integral be replaced by the factor ($2-
{{\cal F}_{\rm over}}$) in the second integral.  Setting ${{\cal F}_{\rm over}}
= 1.0$ clearly recovers the integral equation that applies when viscous
dissipation is neglected altogether (see Roxburgh's paper): this case 
corresponds to the maximum possible amount of overshooting.  The minimum size
of a convective core is obtained when the radiative-convective boundary is
determined from the Schwarzschild criterion.  (In what follows, the
non-overshooting calculations are labelled ``${{\cal F}_{\rm over}} = 0.0$'' even
though, in this special case, the overshooting subroutine is bypassed.)

One goal of this investigation is to determine which value of ${{\cal F}_{\rm
over}}$ leads to the best agreement between theoretical models and the observed
CMD of NGC$\,$7789.  Knowing this will provide a valuable calibration point for
the variation of this parameter with mass (and possibly [Fe/H]), which we hope
to derive from similar considerations of other open clusters that encompass a
wide range in age and metal abundance.  But the distance scale is an important
ingredient in such analyses.  In a highly-reddened cluster like NGC$\,$7789, it is
arguably the best approach to infer the distance modulus from the magnitude of
the clump stars.  Based on published observations for many open clusters,
Twarog et al. (1997) have suggested that ``over the age range from NGC$\,$7789 to 
Mel 66 (approximately 1 to 5 Gyr), the mean $M_V$ is $0.6\pm 0.1$''.  Since
the mean magnitude of the clump in NGC$\,$7789 is $V = 13.0\pm 0.05$ mag (see
Figs.~\ref{pfig7a} and \ref{pfig7b}), the Twarog et al.~$M_V$ estimate implies 
that the apparent distance
modulus of NGC$\,$7789 is close to $12.40$ mag.  The reddening then follows from the
requirement that the predicted zero-age main sequence, which is independent of
whether or not overshooting occurs, match the observed main sequence for the
unevolved stars, i.e., those fainter than $M_V\sim 4$.  (Deriving this
quantity from a main-sequence fit to stellar models should be especially
reliable because the theoretical calculations have been normalized to the Sun:
we assume $M_{V,\odot} = 4.84$, $(B-V)_\odot = 0.64$, and $(V-I)_\odot = 0.72$).

Figures~\ref{pfig10a} and \ref{pfig10b} illustrate fits of evolutionary 
tracks for ${{\cal F}_{\rm over}} = 0.0$ and 1.0, respectively, to the NGC$\,$7789 
CMD assuming $(m-M)_V = 12.40$ and $E(V-I) = 0.38$.  The tracks are the same as
those used by Rosvick \& VandenBerg (1998) in their analysis of NGC 6819 $BV$ photometry, 
though we have extended those grids to include
evolutionary sequences for 1.8 and 1.9$\,{{\cal M}_\odot}$ stars.
It is immediately apparent that neither set of models correctly predicts the
observed TAMS (``terminal age main sequence'').  The predicted TAMS, which is
indicated by the nearly vertical {\it solid} curve connecting the red end of the hook feature,
is either much too blue, in the case of the non-overshooting models, or somewhat 
too red, in the case of the models for ${{\cal F}_{\rm over}} = 1.0$: this is in
comparison with the observed location of the bright end of the cluster main 
sequence.  An intermediate amount of overshooting is clearly indicated and, as
shown in Figure~\ref{pfig11}, models that assume ${{\cal F}_{\rm over}} 
= 0.5$ provide quite a good match to the observed TAMS.

The tracks for masses from 1.1 to 1.9$\,{{\cal M}_\odot}$ were extended to
include the giant-branch phase and, using a modified version of the Bergbusch \&
VandenBerg (1992) interpolation code (see \cite{bv98}), isochrones were produced 
for a suitable range in age.  
Interestingly, the calculations revealed that, for masses $\gta$ 1.8$\,{{\cal M}_\odot}$,
the RGB tip magnitudes decrease rapidly with increasing mass. For instance, the giant-branch
tip for a 1.9$\,{{\cal M}_\odot}$ star is predicted to be $\Delta$$ M_{bol} \approx 0.7$ mag less
luminous than that for a 1.8$\,{{\cal M}_\odot}$ star. This fairly abrupt decrease in the
tip magnitude, and hence in the length of the RGB, signals the approach to the transition mass
between quiescent and violent helium ignition (the so-called ``helium flash'').
In the case of non-overshooting models (see, e.g., Sweigart et al.~1989), this
transition occurs near 2.3$\,{{\cal M}_\odot}$. 
However, it is one of the well-known consequences
of convective overshooting on the main sequence (see Bertelli et al. 1986, and
references therein) that the lower mass limit for quiescent helium ignition is
reduced (to $\approx$ 2.0$\,{{\cal M}_\odot}$ for our ${{\cal F}_{\rm over}} = 0.5$
models). Thus, whether or not an intermediate-age cluster shows an extended
giant branch provides a useful constraint on its distance and age (or on the
importance of convective overshooting), as was first pointed out by Barbaro \& Pigatto (1984).

Indeed, this turns out to be an important consideration in the present analysis.
Figure~\ref{pfig12} illustrates the superposition of a 1.5 Gyr isochrone onto the NGC$\,$7789
photometry, on the assumption of the same distance modulus and reddening that were adopted 
in the previous three figures.  This is clearly not
a ``best-fit'' isochrone as it is readily apparent that one for a slightly
younger age is needed to fit the turnoff observations.  However, besides being
too red, the predicted
giant branch at an age of 1.5 Gyr is not quite as extended as it needs to be
to match the observed RGB, and that for any younger age (higher turnoff mass) 
will be even stubbier.
Thus, in order for our overshooting models to match both 
the turnoff morphology {\it and} the well-developed giant branch possessed by NGC$\,$7789, 
we are forced to conclude that the cluster distance modulus must be less than 
$(m-M)_V = 12.40$.

A more favorable comparison between the solar-abundance models and the NGC$\,$7789
CMD is obtained if $(m-M)_V = 12.20$.  This is illustrated in 
Figure~\ref{pfig13}, which indicates that a value of ${{\cal F}_{\rm over}} = 
0.5$ still provides a good fit to the observed TAMS, and in Figure
\ref{pfig14}, which shows that a 1.6 Gyr isochrone faithfully reproduces the
cluster's main sequence and RGB fiducials.  However, the comparison between
theory and observations is still not completely free of difficulty.  For the
first time in this analysis, we have superposed a theoretical ZAHB locus on the
photometry and its location relative to that of the core He-burning stars seems
problematic.  This ZAHB consists of a sequence of 29 models having the same
helium core mass, ${{\cal M}_c}$, but a range in total mass from 0.50$\,{{\cal M}_\odot}$, 
at the blue end, to 1.8$\,{{\cal M}_\odot}$, at the bright
end past the red ``nose''.  If no mass is lost in the precursor evolutionary
phases, the clump stars should populate the bright end of the ZAHB:
mass-losing stars should be fainter and redder than this point, eventually
becoming bluer again if the mass loss has been severe.  But there are many stars
lying along the extension of the ZAHB to bluer colors and brighter magnitudes.

The same ``anomaly'' was found by Rosvick \& VandenBerg (1998) 
in the case of the $\sim 2.4$ Gyr
open cluster NGC$\,$6819.  They suggested that the brightest of the clump stars
might be the descendants of blue stragglers, given that there is a confirmed
blue straggler population in that cluster.  NGC$\,$7789 also contains
significant numbers of such stars (see Figs.~\ref{pfig7a} and \ref{pfig7b}) 
and one cannot help but speculate
that their progeny might lie along the extension of the ZAHB to higher mass (if
blue stragglers are significantly more massive than the cluster turnoff stars).
However, there appears to be too many excessively bright clump stars for this
to be the entire explanation.

We may, of course, be placing too much reliance on the accuracy of the ZAHB
models.  The reader will notice that the flat part of the ZAHB is very faint.
This is for the reason that the predicted core mass in the overshooting models,
particularly when the total stellar mass approaches the transition mass, is
much lower than the values found in non-overshooting models.  For instance, a
1.7$\,{{\cal M}_\odot}$ star having solar abundances is predicted to have
${{\cal M}_c} = 0.4659 {{\cal M}_\odot}$ if ${{\cal F}_{\rm over}} = 0.0$ (no 
overshooting) versus 0.4358$\,{{\cal M}_\odot}$ if ${{\cal F}_{\rm over}} = 0.5$.
This reduction in the core mass causes a decrease of $\approx 0.3$ mag in the 
luminosity of the ZAHB.  Thus, in all clusters whose upper main-sequence stars
burn hydrogen in convective cores, the theoretical ZAHB locus depends quite 
sensitively on the amount of core overshooting which is assumed to occur in
the main-sequence phase.  This is clearly an important complication in the interpretation
of star cluster CMDs, particularly if ${{\cal F}_{\rm over}}$ varies with mass.  

But, regardless of whether or not the ZAHB in Fig.~\ref{pfig14} is somewhat too faint
(due, perhaps, to a problem with the models), there also seems to be a bit of a
mismatch between the predicted and observed colors of the clump stars.  While it
is always very risky to try to draw conclusions from color information, the 
assumption of a constant reddening may be partly responsible.  As Twarog et 
al.~(1997; see references therein) have noted, when the reddening is 
substantial, a red giant will exhibit a smaller $E(B-V)$ than a hotter 
main-sequence star if both are obscured by the same dust layer.  They suggest 
that the reddening correction applied to the NGC$\,$7789 giants should be about 
0.03 mag smaller than the value applied to its turnoff stars.  Such a 
differential correction would go in the right direction to improve the agreement
between the models and the observed clump (while worsening slightly the fit to 
the brightest giants).

Reasonably good fits of isochrones to the NGC$\,$7789 CMD can be obtained
on the assumption of smaller distances, with consequent modest increases in the
inferred age --- e.g., 1.9 Gyr if $(m-M)_V = 11.9$ --- but, if the apparent
distance modulus is less than 12.0, then the predicted giant branch is well to 
the blue of its observational counterpart.  The main argument that we can offer
against the possibility that such discrepancies are due to inadequacies in the
adopted color transformations, is that the same models produce a superb match
to the CMD of NGC$\,$6819 (see Rosvick \& VandenBerg 1998), which is only 0.6--0.9
Gyr older than NGC$\,$7789.  Pending further developments, we therefore believe
that Fig.~\ref{pfig14} portrays the best interpretation of the NGC$\,$7789 photometry {\it
if} this cluster has the same chemical composition as the Sun.

Two additional points can be made in support of the basic cluster parameters that
we have derived. First, given that $E(V-I) = 1.25E(B-V)$ (\cite{bb88}), the reddening 
value that was inferred from a main-sequence fit to stellar models---namely
$E(V-I) = 0.35$---corresponds to $E(B-V) = 0.28$. This is very close to the mean of the 
$E(B-V)$ values listed in Table~\ref{redmet}. And second, if we adopt 
$A_{V} = 3.1E(B-V)$ and $A_{I}=1.98E(B-V)$ (\cite{b98}), then the adopted apparent
distance modulus in $V$, i.e., $(m-M)_{V}=12.20$, translates to $(m-M)_{I}=11.88$.
As illustrated in Figure~\ref{pfig15}, the quality of the resultant isochrone fit
on the [$M_{I}, (V-I)_{0}$]-plane is quite comparable to that obtained on the 
[$M_{V}, (V-I)_{0}$]-plane (Fig.~\ref{pfig14}). But the main point to be noted
here is that the mean clump magnitude is essentially identical with the value
M$_{I}$ = $-$0.23 $\pm$ 0.03 that is obtained from a volume-limited sample of
more than 200 clump stars observed with {\it Hipparcos} (see, e.g.
\cite{sg98}).

We now turn to the possibility that it has [Fe/H] $= -0.2$, which is more 
consistent with the available metallicity determinations.  By adopting various
choices for the apparent distance modulus and examining the best-fit isochrone
for that distance, it quickly became evident that $(m-M)_V$ had to be $\lta 
12.15$ in order for the giant branch of the corresponding isochrone to be at
least as long as the observed one.  The isochrone fit that we obtained to the
NGC$\,$7789 photometry on the assumption of this distance upper-limit is shown in 
Figure~\ref{pfig16}: the inferred age is 1.7 Gyr.  Except for the giant 
branch, the fit is quite comparable to that given in Fig.~\ref{pfig14}.  
If anything, the predicted ZAHB magnitudes agree slightly better with 
those of the observed clump, though a modest color offset (0.05 mag, say) remains.  
As before, a value of ${{\cal F}_{\rm over}} = 0.5$ yields good agreement 
between the predicted and
observed TAMS.  (This is sufficiently clear in Fig.~\ref{pfig16} that we decided not to
include a plot, like Fig.~\ref{pfig13}, which compares evolutionary tracks with the
observed CMD.)
 
Because of the giant branch discrepancy, it is tempting to conclude that
[Fe/H] $= -0.2$ is too low.  Obviously, much more work is needed to better 
establish the reliability of the models before such an inference could be taken
seriously, though, in support of this possibility, we note that Twarog et 
al.~(1997) recently derived [Fe/H] $= -0.08$ from DDO photometry.  What is
fortunate is that the derived value of ${{\cal F}_{\rm over}}$ as well as the
inferred distance and age are only weakly dependent on the cluster metallicity. 
Thus, regardless of which value of [Fe/H], in the range $-0.2\le$ [Fe/H] $\le 
0.0$, applies to NGC$\,$7789, it seems clear that the models must assume a value of
${{\cal F}_{\rm over}}$ close to 0.5 in order to match the observed CMD in the 
vicinity of the turnoff, that the cluster distance is approximately $(m-M)_V = 
12.2$, and that the age of NGC$\,$7789 is 1.6 -- 1.7 Gyr.

\section{Conclusion} \label{CON}

We have presented extensive CCD photometry in $V$ and $I$ for the open cluster
NGC$\,$7789 that establishes its fiducial sequences on the $(V,V-I)$--plane down
to $V\sim 21$ ($M_V\sim 9$).  Despite the relatively high $E(V-I) \approx
0.35$, these sequences are very tight and well-defined, indicating that the
reddening is nearly constant across the face of the cluster.  A confrontation of
these data with modern stellar evolutionary models has demonstrated, beyond
any doubt, that significant convective core overshooting must occur in the main
sequence phase.  Non-overshooting models are unable to match either the 
observed morphology in the vicinity of the turnoff or the location of the
TAMS relative to that of the main sequence.  Perhaps the most novel aspect of
our analysis is that the length of the red-giant branch has been used to set a
lower limit to the cluster age (1.6 Gyr) and an upper limit to its 
distance [$(m-M)_V = 12.2$].  An important, well-known consequence of convective
core overshooting is to reduce the maximum mass of stars that undergo the
helium flash: above this limit, helium-burning is ignited quiescently and
extended giant branches are not produced.  As NGC$\,$7789 has a well-developed,
but not fully extended RGB, its giants must be just slightly below the lower
mass limit where non-violent ignition of helium takes place.
 
This paper is the second in a loosely-connected series (the first was the
study of NGC$\,$6819 by Rosvick \& VandenBerg 1998) to calibrate the
dependence of the extent of overshooting on turnoff mass (and possibly metal
abundance). As soon as that calibration is completed, a new grid of evolutionary 
tracks and isochrones will be computed (by D.A.V.) for application to intermediate-age
stellar populations\footnote{In the meantime, the models used in both this investigation
and that by Rosvick \& VandenBerg may be obtained by e-mailing a request for them to D.A.V.}.

\acknowledgments

We thank an anonymous refree for a thoughtful and helpful report and Pat Dowler 
for making available the overshooting subroutine that he
developed as part of his M.Sc.~thesis research.  NSERC is acknowledged for a
grant to P.B.S.~and J.E.H.~that provided support for M.G.  This work was also 
partially supported by an NSERC Operating Grant to D.A.V.

\clearpage

\begin{figure}
\figurenum{1a}
\caption[]
{The area observed in the vicinity of NGC$\,$7789. 
Note the approximate magnitude scale in the lower right-hand corner.
Units on the horizontal and vertical axes are pixels, where 2048 pixels 
correspond to 9$\farcm$2. 
The center of the panel (4628, 3873) is nearly identical with the cluster center. 
}
\label{pfig1a}
\end{figure}

\begin{figure}
\figurenum{1b}
\caption[]
{
Same as Fig.~\ref{pfig1a}, but indicates stars which are believed to be members (see section 3).
}
\label{pfig1b}
\end{figure}

\begin{figure}
\figurenum{2}
\caption[]
{Residuals (in the sense of ``present'' $-$ ``standard'') in $V$ and $V~-~I$
as functions of apparent magnitude and color for the primary standard stars.
{\it Crosses}: Landolt (1992) standards; {\it pluses}: Landolt stars (\cite{s97}); 
{\it open squares}: M$\,$67 stars (\cite{mmj93}); {\it solid circles}: NGC$\,$7006 stars
(\cite{dav92}); and {\it open circles}: NGC$\,$7006 stars (\cite{s97}).} 
\label{pfig2}
\end{figure}

\begin{figure}
\figurenum{3a}
\caption[]
{$V$ magnitude comparison between our data and several published data sets 
obtained with photoelectric photometry. {\it Open circles}: Burbidge \& Sandage (1958),
{\it crosses}: Breger (1982), {\it squares}: Janes (1977), 
~{\it solid triangles}: Coleman (1982), ~{\it open triangles}: 
Jennens \& Helfer (1975), and {\it solid circles}: Breger \& Wheeler (1980). 
Stars with large differences in the plot are M589 (one from \cite{j77} and 
the other from \cite{jh75}), M467 (\cite{b82}) and M789 (\cite{b82}).  
In addition, M292 (\cite{b82}) and M864 (\cite{j77}) have
$|\Delta{V}| > 0.5$. $V$ and $V-I$ on the horizontal axes  are the magnitude 
and color from the present study.
See Table \ref{Tcomp} for a statistical summary of these plots.
}
\label{pfig3a}
\end{figure}

\begin{figure}
\figurenum{3b}
\caption[]
{Same as Fig. \ref{pfig3a}, but comparing the present data with 
published photographic photometry. 
{\it Crosses}: Burbidge \& Sandage (1958), {\it open circles}: McNamara (1980).  
Four stars (M589, M1012, M818, and M717) from Burbidge \& Sandage (1958) 
and three stars (M589, M1012, and M1054) from McNamara (1980) have
$|\Delta{V}| > 0.5$. 
}
\label{pfig3b}
\end{figure}

\begin{figure}
\figurenum{3c}
\caption[]
{Same as Fig. \ref{pfig3a}, but comparing the present data with published
CCD photometry
(Jahn et al. 1995). The total number of stars in this plot is 3030 
including two stars with $|\Delta{V}| > 0.5$.
}
\label{pfig3c}
\end{figure}

\begin{figure}
\figurenum{4}
\caption[]
{
CMD for 15617 stars which were detected more than once in both V and I frames and
also have $\sigma_{(V-I)}< 0.1$. 
}
\label{pfig4}
\end{figure}

\begin{figure}
\figurenum{5}
\caption[]
{
CMD for 5915 stars which are not used for the analysis because 
$\Delta (V-I) \ge 0.1$ or they were detected only once in either of the 
$V$ or $I$ frames. The solid line is the fiducial MS line which is obtained by eye
fitting through the mean locus in Fig.~\ref{pfig4}.
}
\label{pfig5}
\end{figure}

\begin{figure}
\figurenum{6}
\caption[]
{
The standard errors in V plotted as a function of $V$ magnitude.
}
\label{pfig6}
\end{figure}

\begin{figure}
\figurenum{7a}
\caption[]
{A CMD which incorporates membership information for giant branch, blue straggler 
and upper-main sequence stars, as follows: {\it plus signs}: proper motion members 
(\cite{ms81}); {\it open circles} and {\it ``V''}: radial velocity members including
velocity variables (\cite{ghms98}); {\it open squares}: blue stragglers (see Table~\ref{BSMem}).
{\it Dots} are stars for which no membership information other than their location
in the CMD is available.
}
\label{pfig7a}
\end{figure}

\begin{figure}
\figurenum{7b}
\caption[]
{
Same as Fig.~\ref{pfig7a}, but for only those stars believed to be members.
The clump star region is shown in the inset box with the expanded scale.
}
\label{pfig7b}
\end{figure}

\begin{figure}
\figurenum{8}
\caption[]
{CMD of M67 (\cite{mmj93}) overplotted by the fiducial sequence for the NGC$\,$7789
MS and its clump ({\it large cross}), which are shifted by 
$\Delta${($V-I$)}~=~0.25 and $\Delta{V}$~=~2.45. 
As shown in the inset box, these adjustments produce an approximate centering of 
the NGC$\,$7789 clump onto that for M$\,$67. The small {\it filled circles} represent
the observed clump stars in M$\,$67 while all other symbols represent NGC$\,$7789 
clump stars (see the previous figure).
} 
\label{pfig8}
\end{figure}

\begin{figure}
\figurenum{9}
\caption[]
{As Fig. \ref{pfig8}, but NGC$\,$7789's data are shifted by 
$\Delta${(V$-$I)}~=~0.29 and $\Delta{V}$~=~2.62. As shown in the inset box,
these offsets cause the M$\,$67 clump stars to be redder and fainter, in the mean,
than their counterparts in NGC$\,$7789. This is approximately consistent with the
expectations from theoretical ZAHB models for 1.2 and 1.6$\,{\cal M}_{\odot}$ 
(which assumes some mass loss).
}
\label{pfig9}
\end{figure}

\begin{figure}
\figurenum{10a}
\caption[]
{
Superposition of non-overshooting evolutionary tracks for 0.8 -- 1.9$\,{{\cal M}_\odot}$
onto the NGC$\,$7789 CMD. (To highlight the comparison between theory and observations at the
upper end of the main sequence, the plot has been restricted to M$_{V}$ $\leq$ 5: subsequent
figures illustrate fits of models to the entire CMD.) The adopted reddening and distance modulus
are as noted. The ZAMS and TAMS points on the tracks are connected by {\it thick solid curves}.
The upper main-sequence population in NGC$\,$7789 extends as far as the {\it dashed line}, 
which indicates our estimate of the location of the observed TAMS. Because the predicted and 
observed TAMS loci do not coincide, stellar models that neglect covective core overshooting 
are deemed inappropriate.
}
\label{pfig10a}
\end{figure}

\begin{figure}
\figurenum{10b}
\caption[]
{As Fig. \ref{pfig10a}, but for models with the maximum possible
extent of overshooting according to the Roxburgh (1978, 1989)
criterion. For the sake of clarity the dashed line in the previous figure has not
been reproduced but it is obvious that the predicted TAMS again fails to match
observed one, indicating (in this case) that too much convective overshooting has been assumed.
}
\label{pfig10b}
\end{figure}

\clearpage
\begin{figure}
\figurenum{11}
\caption[]
{As Fig. \ref{pfig10b}, but for models with an intermediate 
amount of overshooting according to the parameterized version of Roxburgh (1978, 1989)
criterion that we have adopted. The predicted TAMS now reproduces the observed one quite well.
}
\label{pfig11}
\end{figure}

\begin{figure}
\figurenum{12}
\caption[]
{The CMD for NGC$\,$7789 with an isochrone for 1.5 Gyr superimposed. 
While the match between theory and observations is generally good, 
the theoretical giant branch does not extend far enough to fit
the observed red-giant branch. 
}
\label{pfig12}
\end{figure}

\begin{figure}
\figurenum{13}
\caption[]
{NGC$\,$7789 CMD on which evolutionary tracks for
0.7$-$1.9$\,{\cal M}_{\odot}$ are superimposed. The adopted reddening and
distance modulus are as noted. ZAMS and TAMS points are connected by 
thick solid curves. The models are calculated with an intermediate amount of
overshooting and the TAMS agrees well with the observed one. 
}
\label{pfig13}
\end{figure}

\begin{figure}
\figurenum{14}
\caption[]
{CMD for NGC$\,$7789 with an isochrone for 1.6 Gyr superimposed, along
with the corresponding theoretical ZAHB.
}
\label{pfig14}
\end{figure}

\begin{figure}
\figurenum{15}
\caption[]
{$M_{I}$ versus $V-I$ CMD for NGC$\,$7789 with an isochrone for 1.6 Gyr superimposed, along
with the corresponding theoretical ZAHB.
}
\label{pfig15}
\end{figure}

\begin{figure}
\figurenum{16}
\caption[]
{CMD for NGC$\,$7789 with a 1.7 Gyr isochrone for [Fe/H] $= -0.2$ superimposed.
}
\label{pfig16}
\end{figure}

\begin{table}
\dummytable\label{redmet}
\end{table}

\begin{table}
\dummytable\label{coefFix}
\end{table}

\begin{table}
\dummytable\label{coefCol}
\end{table}

\begin{table}
\dummytable\label{Tcomp}
\end{table}

\begin{table}
\dummytable\label{BSMem}
\end{table}

\end{document}